\documentclass[ aps,prl, twocolumn, superscriptaddress,letterpaper]{revtex4}
\usepackage{graphicx}
\usepackage{amssymb,amsmath}
\usepackage{float}
\begin{document}

\title{Coupling the valley degree of freedom to antiferromagnetic order} 

\author{Xiao Li}
\affiliation{International Center for Quantum Materials, Peking University, Beijing 100871, China.}
\author{Ting Cao}
\altaffiliation{Present address: Department of Physics, University of California at Berkeley, Berkeley, CA 94720, USA}
\affiliation{International Center for Quantum Materials, Peking University, Beijing 100871, China.}
\author{Qian Niu}
\affiliation{International Center for Quantum Materials, Peking University, Beijing 100871, China.}
\affiliation{Department of Physics, University of Texas at Austin, Austin, TX 78712, USA. }
\author{Junren Shi}
\author{Ji Feng}
\email{jfeng11@pku.edu.cn}
\affiliation{International Center for Quantum Materials, Peking University, Beijing 100871, China.}

\begin{abstract}

Conventional electronics are based invariably on the intrinsic degrees of freedom of an electron, namely, its charge and spin. The exploration of novel electronic degrees of freedom has important implications in both basic quantum physics and advanced information technology. Valley as a new electronic degree of freedom has received considerable attention in recent years. In this paper, we develop the theory of spin and valley physics of an antiferromagnetic honeycomb lattice. We show that by coupling the valley degree of freedom to antiferromagnetic order, there is an emergent electronic degree of freedom characterized by the product of spin and valley indices, which leads to spin-valley dependent optical selection rule and Berry curvature-induced topological quantum transport. These properties will enable optical polarization in the spin-valley space, and electrical detection/manipulation through the induced spin, valley and charge fluxes. The domain walls of an antiferromagnetic honeycomb lattice harbors valley-protected edge states that support spin-dependent transport. Finally, we employ first principles calculations to show that the proposed optoelectronic properties can be realized in antiferromagnetic manganese chalcogenophosphates (MnPX$_3$, X = S, Se) in monolayer form.

\end{abstract}

\maketitle

The exploration of novel electronic degrees of freedom \cite{Rycerz07, Gunawan06, Takashina06, Eng07} has been a fairly important topic recently, as the future of information technology is likely to be levered upon it. The intrinsic degrees of freedom of an electron, namely, its charge and spin, has been the basis for the society-transforming information technologies, i.e., electronics and spintronics. Additional electronic degree of freedom, if present, will offer immense potential for information encoding and manipulation at the microscopic level.  The notion of valleytronics \cite{Rycerz07} on honeycomb lattices  has received considerable attention in recent years.  \cite{Xiao07, Yao08, Cao12, Xiao12, Mak12, Zeng12} When the centrosymmetry of the honeycomb lattice is broken, such as in gapped graphene, there arise an inequivalent, degenerate pair of valleys in the momentum-space electronic structure. The valley excitations, protected by the suppression of intervalley scattering, have contrasting optical and transport properties ensured by quantal helicity.\cite{Xiao07, Yao08} In the most recent experimental progress in monolayer group VI transition metal dichalcogenides, the identity of valleys manifests as valley-selective circular dichroism, leading to substantial valley polarization with circularly polarized light, offering a potential arena to the eventual realization of valleytronics. \cite{Cao12, Mak12, Zeng12}

Central to these endeavors are two tasks, namely, developing theoretical paradigms and subsequent materials discovery, the latter of which makes possible experimental measurements that put the theory \cite{Xiao07, Yao08, Cao12} to test \cite{Cao12, Mak12, Zeng12}. Evidently, it is important to broaden the choice of materials, beyond transition metal dichalgogenides, \cite{Cao12, Xiao12, Mak12, Zeng12} with which novel degrees of freedom (beyond charge and spin) of Bloch electrons can be accessed. In previous theoretical and experimental developments, attention has been paid to the absence of inversion center in the lattice space group. Here we show that the pseudospin symmetry present in the initially symmetric honeycomb lattice hints at a nontrivial transformation, which leads to an emergent degree of freedom characterized by the product of spin and valley indices. We further propose that such a Fermionic system may be observed on a bipartite honeycomb lattice that assumes the N$\acute{\text{e}}$el antiferromagnetism (afm), which can possess chiral electronic excitation concomitant to the spin density wave, as well as spin-dependent transport properties on the edge states. In the last part of the paper, we employ first principles calculations to show that the proposed physics can be realized in antiferromagnetic manganese chalcogenophosphates (MnPX$_3$, X = S, Se) in monolayer form. We also analyze the general consequence of symmetry in the afm Fermionic honeycomb lattice. A potential transition to the topologically non-trivial quantum spin Hall state is briefly discussed.

Valley as an electronic degree of freedom has been suggested in a spinless Fermion model on a honeycomb lattice with a symmetry-breaking perturbation to the sublattices.\cite{Xiao07,Yao08} A honeycomb lattice can be defined as a three-connected two-dimensional net, with the connection vectors for bond length $b$ pointing toward three nearest neighbors, ${\bf d}_{1,2} = ({\pm\sqrt{3}\hat{\bf x}+\hat{\bf y}})b/2,$ and ${\bf d}_3 = -b \hat{\bf y}$. The corresponding lattice constant $a=\sqrt{3}b$. The two sublattices of a honeycomb structure is conventionally denoted A and B, respectively, which correspond to a binary degree of freedom called isospin. The low-energy quasiparticle states at ${\bf K}_{\pm}=\pm 4\pi/3a\hat{\bf x}$ are assigned a valley index $\tau=\pm1$, indicating the valley contrasting physics. It is tied to physical measurables, such as the orbital magnetic moment, $\mathcal{M}({\bf K}_{\pm})=\tau\mu_{\text{B}}^{*}$, where $\mu_{\text{B}}^{*}$ is the effective Bohr magneton.\cite{Xiao07} Concomitant to the orbital magnetic moment is non-vanishing Berry curvatures,\cite{Xiao10} $\Omega({\bf K}_\pm)=\hbar \mathcal{M}({\bf K}_\pm)/e\varepsilon({\bf K}_\pm)$, where $e$ and $\varepsilon({\bf K}_\pm)$ are the electronic charge and the band energy at ${\bf K}_\pm$ , respectively.\cite{Yao08}

When the spin degree of freedom ($s=\pm1/2$) is taken into account, spontaneous symmetry breaking becomes viable in the dichromatic Shubnikov group. Imposing dichromatic coloring on the spin and lattice, we propose a spin-full Hamiltonian for the low-energy quasiparticles near ${\bf K}_\pm$ of a honeycomb lattice,
\begin{equation}
\mathcal{H}_{{\bf }}^{(s\tau)}=v_{\text{F}}s_{0}(\tau_{z}\sigma_{x}p_{x}+\tau_{0}\sigma_{y}p_{y})+ms_{z}\tau_{0}\sigma_{z},\label{qg1}
\end{equation}

\noindent where ${\bf p}$ is the momentum and $v_{\text{F}}$ is the massless Fermi velocity. Here, $\tau_{\alpha}$, $\sigma_{\alpha}$ and $s_{\alpha}$ $(\alpha=x,y,z,0)$ are the Pauli matrices for the valley, isospin and spin degrees of freedom, respectively. The mass term, $m$, corresponding to a symmetry-breaking perturbation, admits a band gap $\Delta=2m$ for both spins. With the spin-dependent Hamiltonian, $\mathcal{M}({\bf K}_{\pm})=2s\tau\mu_{\text{B}}^{*}$. Therefore, the product of spin and valley indices identifies a new degree of freedom of electrons, which we shall call the coupled spin-valley, $s\cdot\tau$, degree of freedom. The $s\cdot\tau$ index bears the virtue of a good quantum number, for so long as inter-valley scattering is suppressed it mimics the two spin states of an electron. 

\begin{figure}[H]
\centering{}\includegraphics[width=6.5cm]{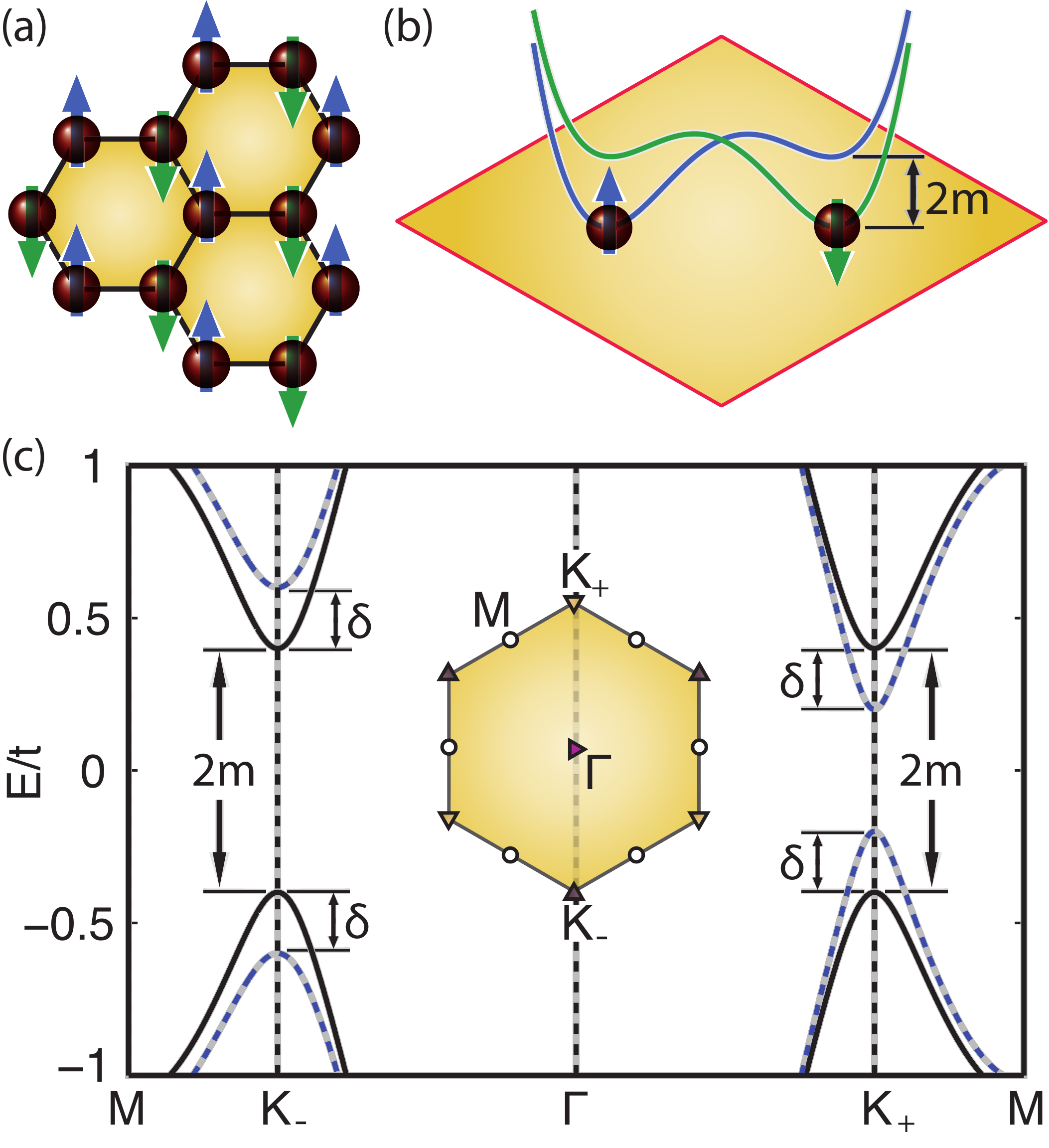}
\caption{The antiferromagnetic honeycomb lattice. (a) N$\acute{\text{e}}$el antiferromagnetism of a honeycomb lattice. Up and down arrows indicate a staggered spin-density wave. (b) A schematic of the spin-dependent lattice potential corresponding to the afm order. (c) Low-energy quasiparticle bands of a NN hopping afm Hamiltonian,  Solid lines assume zero spin-valley coupling, while dashed lines take into account the spin-valley coupling. Inset: the Brillouin zone and high symmetry points.
 }
\end{figure}

As the spinless Fermion model describes a non-magnetic system, \cite{Xiao07} the dichromatic Hamiltonian of equation \eqref{qg1} represents an antiferromagnetic system. Inspection of Hamiltonian \eqref{qg1} reveals that the $s\cdot\tau$ degree of freedom arises upon engaging the electrons with the spin-dependent on-site Hubbard potential, $Un_{j}^{(s)}n_{j}^{(-s)}$, which characterizes the Coulomb interaction of opposite spins on site $j=$ A, B. \cite{Meng10} Within the Hartree-Fock approximation, the on-site Coulomb energy of two spins can be decoupled as $m^{(s)}n_{j}^{(s)}+m^{(-s)}n_{j}^{(-s)}$, upon an immaterial energy shift. We define a spin-dependent mass term, 
\begin{equation}
m^{(s)} = 2sm,
\end{equation} 
where 
\begin{equation}
m = \frac{U}{2}|\langle n_{j}^{(-s)}\rangle-\langle n_{j}^{(s)}\rangle|,
\end{equation}
characteristic of an afm spin-density wave (Fig. 1a). It is now apparent that with the afm order, each spin sees different potentials on the two sublattices, as represented by $m^{(s)}$ (Fig. 1b) and described by equation \eqref{qg1}.

To pinpoint the essential physics of spin and valley on an afm honeycomb lattice, we analyze a tight-binding Hamiltonian invoking the spin-dependent mass term above and a nearest-neighbor (NN) hopping, $t$,
\begin{equation}
\mathcal{H}_{{\bf k}}=t_{{\bf k}}'s_{0}\sigma_{x}+t_{{\bf k}}''s_{0}\sigma_{y}+ms_{z}\sigma_{z},
\end{equation}
which in the neighborhood of ${\bf K}_{\pm}$ can be linearized to equation \eqref{qg1}. The NN hopping is given as 
\begin{equation*}
t_{{\bf k}}=\sum_{j=1}^{3}-t\exp(-i{\bf k}\cdot{\bf d}_{j})\equiv t_{{\bf k}}'+it_{{\bf k}}'',
\end{equation*} 
where ${\bf d}_{j}$ are the vectors pointing toward the three NN's. The band structure is shown in Fig. 1c. The bands are spin degenerate at each ${\bf k}$-point. Band gaps $\Delta=2m$ are indeed opened at the valleys. An ad hoc spin-valley coupling may be introduced with the parameter $\delta$, as $\mathcal{H}_{{\bf k}}^{(\text{SO})}=\delta s_{z}\tau_{z}\sigma_{z}$. The spin-valley coupling preserves the spin degeneracy, but leads to a renormalization of the valley gaps. The band gaps become enlarged at one valley and reduced at the other, that is, $\Delta=2(m-\tau\delta)$. The fact that spins remain degenerate in this system has to do with the invariance of the Hamiltonian under simultaneous time reversal ($\hat{\mathcal T}$) and spatial inversion ($\hat{\mathcal P}$), although neither $\hat{\mathcal T}$ nor $\hat{\mathcal P}$ alone commutes with the Hamiltonian. We shall revisit this symmetry, $\hat{\mathcal{O}} \equiv \hat{\mathcal P} \hat{\mathcal T}$, later in the paper.

 A few interesting experiments become immediately compelling, to probe and manipulate the $s\cdot\tau$ degree of freedom. When the spin-valley coupling is absent or weak, the two valleys can be considered degenerate (Fig. 2a). The system will have $s\cdot\tau$-selective circular dichroism (CD), similar to gapped graphene\cite{Yao08} and monolayer MoS$_2$ \cite{Xiao12,Cao12,Mak12,Zeng12}. However, because the two spins adopt opposite masses, the optical selection rule at ${\bf K}_{\tau}$ becomes $\eta(s,\tau)=2s\tau$, where $\eta$ is the degree of circular polarization \cite{Yao08,Cao12}. For example, when left-polarized light ($\sigma^{+}$) illuminates the sample, spin up electrons are excited to the conduction band at ${\bf K}_{+}$, while spin-down electrons at ${\bf K}_{-}$, as illustrated in Fig. 2a. $ $When a right-polarized photon is absorbed, it excites down-spins of ${\bf K}_{+}$ and an up spins of ${\bf K}_{-}$.

\begin{figure}[ht]
\centering{}\includegraphics[width=6.5cm]{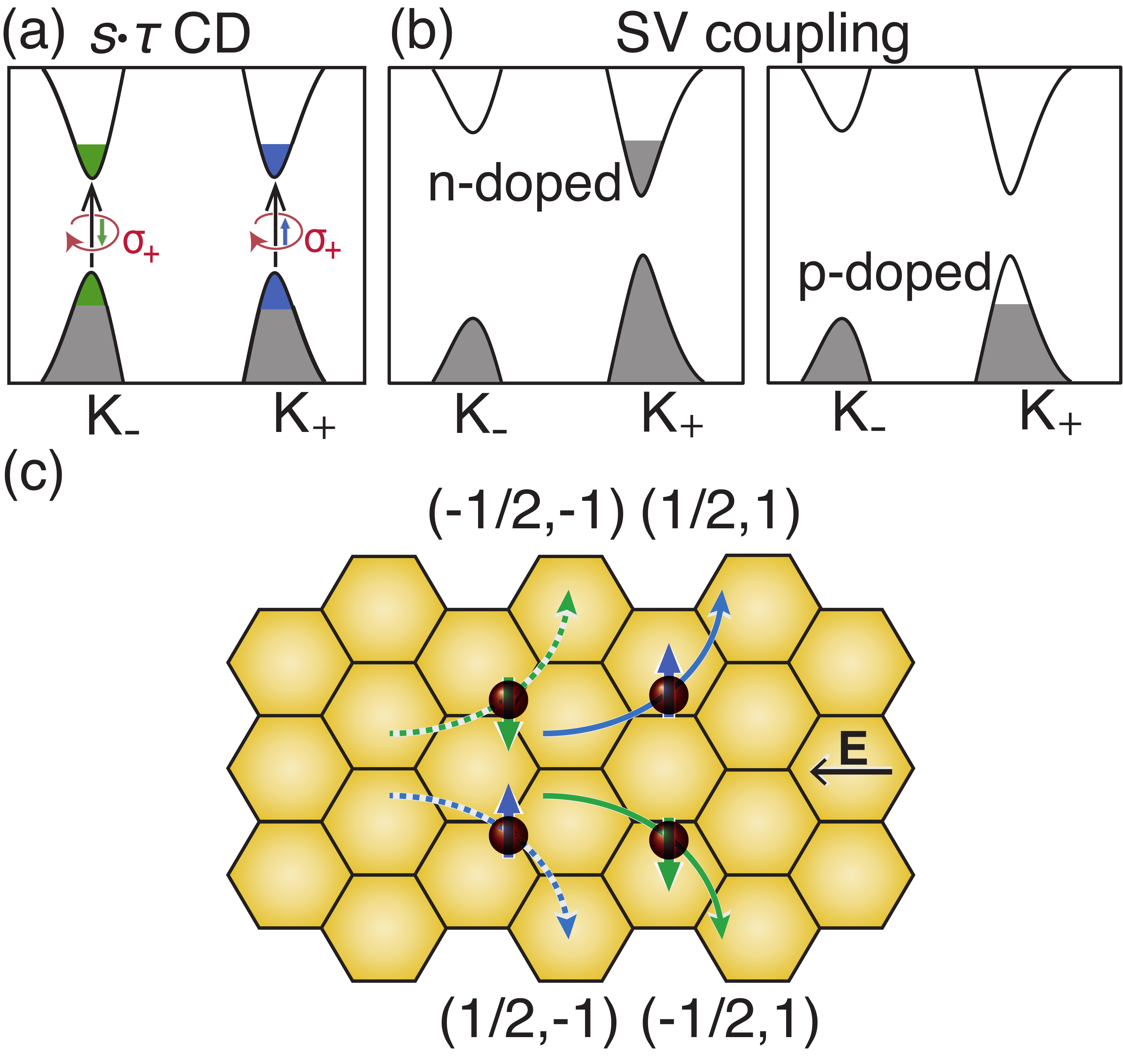} 
\caption{Spin and valley physics of an afm honeycomb lattice. (a) $s\cdot\tau$-selective circular dichroism, in the absence of spin-orbit coupling. (b) When spin-valley coupling is present the valleys can be doped asymmetrically. (c) Electron spin (bottom of conduction bands) fluxes under the action of Berry curvature of the Bloch bands and in-plane electric field. The spin-up and spin-down currents show in blue and green, respectively. Solid and dashed lines stand for the currents from ${\bf K}_{+}$ and ${\bf K}_{-}$, respectively. The spin and valley, $(s,\tau)$, indices are indicated in the parentheses. ${\bf E}$ is an applied in-plane electric field.
 }
\end{figure}

Berry curvature also induces topological quantum transport that allows for electric detection of $s\cdot\tau$ polarization, via the anomalous velocity of Bloch electrons, ${\bf v}_a \sim {\bf E}\times \Omega_{\bf k}$. \cite{Xiao10} In the above dynamical polarization of charge carriers, the spin or spin holes under the action of Berry curvature, $\Omega({\bf k},s)$, of the Bloch bands will exhibit circular dichroic Hall effect (CDHE). Because of the relation with orbital magnetic moments, both Berry curvature and transversal conductivity ($\sigma_{xy}$) depend on $s\cdot\tau$. Charge carriers with $s\cdot\tau=\pm1/2$ (excitable by left- and right-polarized lights, respectively) will have opposite transversal conductivity, as shown in Fig. 2c. Therefore, CDHE is a non-equilibrium charge Hall effect in the presence of a circularly polarized radiation field. 

If we dope the system with electrons or holes at equilibrium, then an applied in-plane electric field will drive a transversal ``valley'' current. Spin currents from all $s\cdot\tau=\pm1/2$ contribute to the transversal transport, resulting in a net accumulation of valley moments and orbital magnetization at the upper and lower edges of the sample (Fig. 2c) with zero transverse charge current. This is the valley Hall effect. In the case of strong spin-valley coupling, the gaps at ${\bf K}_{\pm}$ become different, and thus have different levels of doping. Suppose we dope the system, say, at ${\bf K}_{+}$ with electrons (Fig. 2b). Under an in-plane electric field, the carriers will produce a net transversal spin current without charge current, that is, the pair of currents with $(s,\tau)=(\pm1/2,1)$ in Fig. 2c. This is the anomalous spin Hall effect (SHE). When we switch from n-doping to p-doping (Fig. 2b), the SHE will be characterized by a transversal spin current in the opposite direction, arising from spin holes.

\begin{figure}[h]
\centering{}\includegraphics[width=6cm]{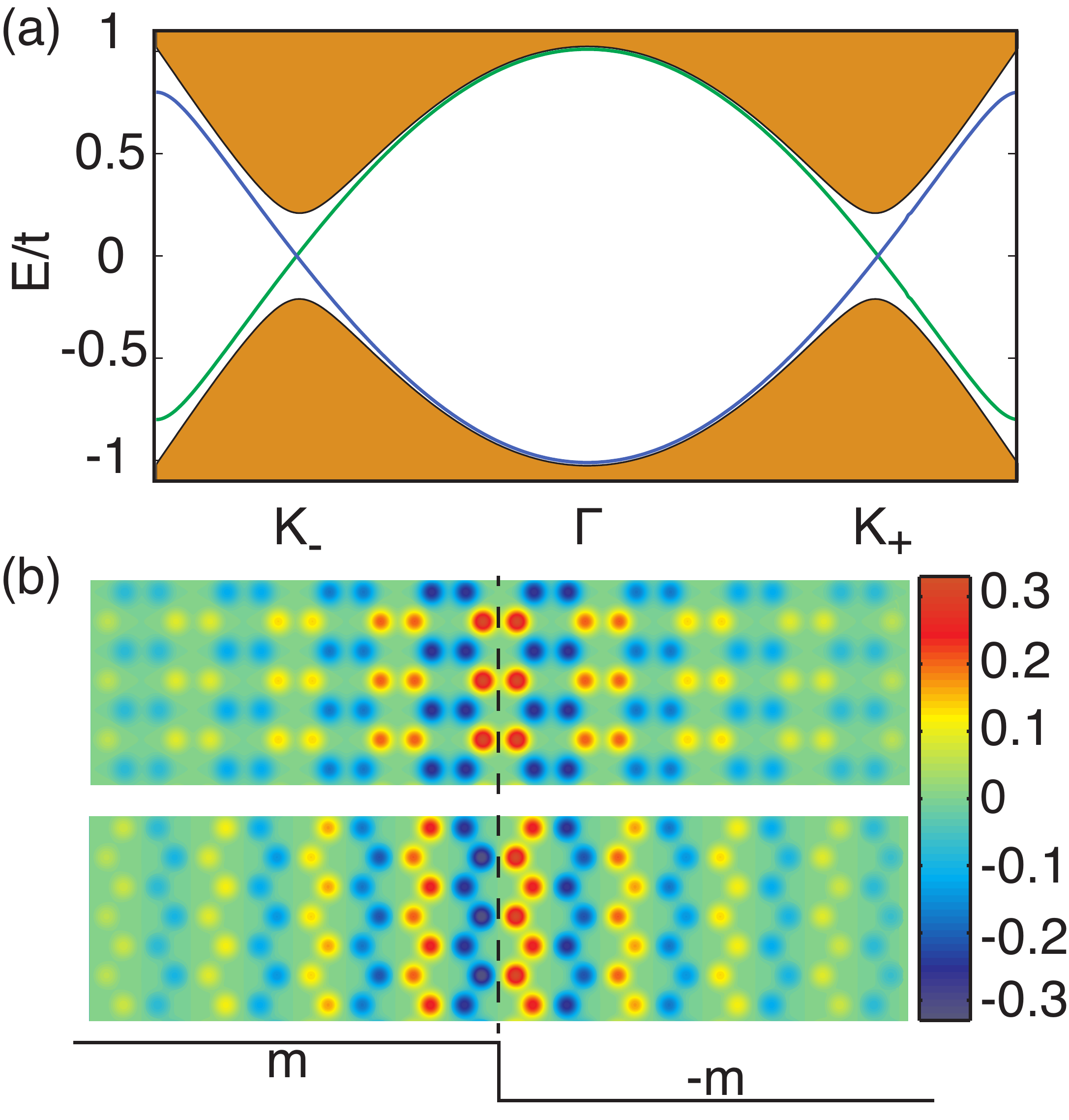}
\caption{Spin-polarized edge states. (a) The band structure in the presence of a zigzag magnetic domain on the afm honeycomb lattice, derived from NN hopping Hamiltonian. A model of a zigzag domain wall is used, which extends 50 unit cells on both sides. The bulk states are lumped into shaded blobs. The mass is set to $m=t/2$. The edge states of two spins are shown in blue and green. (b) The wavefunctions of two spins at the same Dirac points. The amplitudes are convolved on 2-dimensional Gaussians centered on lattice sites for visualization (scale bar in arbitrary units).
}
\end{figure}

Also to be assessed is the topological domain walls that occur naturally with the afm order, of which the band structure is shown in Fig. 3a. There are two spin-polarized bands that arise at the domain wall and intersect to yield a pair of Dirac points, while the bulk states are always spin degenerate. The spin and momentum are clearly locked at individual Dirac point. Unlike the edge states in a topological insulator, these edges states cannot offer the spin-selective channels. However, if a spin is injected into the edge with a prescribed direction of momentum, it can migrate ballistically, enjoying the suppression of back scattering offered by valley protection. It is also of interest to note that the wavefunctions (Fig. 3b) at the same Dirac points have opposite parity for opposite spins. The odd and even parity boundary state may be utilized as spin and/or valley filter, as linear defect in graphene. \cite{Gunlycke11}

Taking one step from the above theoretical exposition, here we suggest actual materials in which the proposed  spin and valley physics can be observed. The selection of materials should meet a few criteria. The candidate materials must 1) have characteristic afm order on a honeycomb lattice; and 2) be a semiconductor, with direct band gaps at high-symmetry ${\bf K}_{+}$, ${\bf K}_{-}$; and 3) have interband transition in the neighborhood of valleys that exhibit circular dichroism. 

These criteria then lead us to manganese chalcogenophosphates, MnPX$_3$, X = S, Se \cite{Ressouche10} in monolayer form. Manganese chalcogenophosphates are layered crystalline materials, in which the interlayer coupling is the relatively weak van der Waals interactions. In principle, all van der Waals bonded layered compounds could be thinned to the monolayer limit by the micromechanical exfoliation technique. \cite{Noveselov04,Novoselov05} Hence, monolayer MnPX$_3$ in all likelihood can be produced. As shown in Fig. 4a, each unit cell in the monolayer MnPX$_3$ is composed of two Mn$^{2+}$ and one {[}P$_2$X$_6${]}$^{4-}$ cluster, the latter isostructural and isoelectronic to molecular ethane (C$_2$H$_6$) in the staggered conformation. The most crucial feature of these compounds is that each Mn\textsuperscript{2+} ions, assuming an $S=\frac{5}{2}$ high-spin state, is antiferromagnetically coupled to its three NN's, forming the afm honeycomb lattice.

\begin{figure}[h]
\centering{}\includegraphics[width=6.5cm]{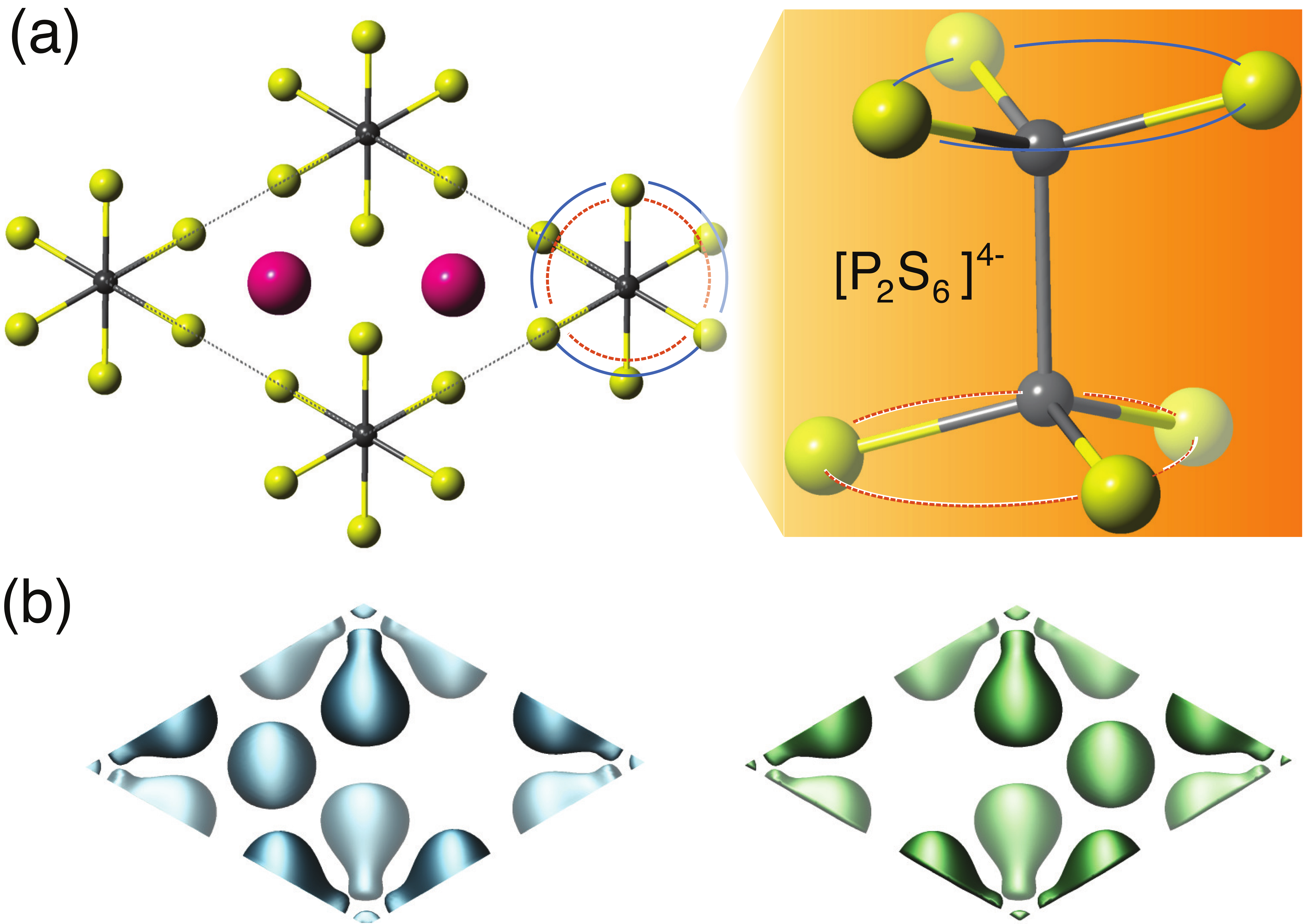}
\caption{
Structure and antiferromagnetism of monolayer MnPX$_3$.  (a) structure showing the unit cell. Purple spheres are Mn, yellow X, and gray P. Some of the computed bond lengths are: P-P = 2.22 \r{A}, P-S = 2.04 \r{A} for MnPS$_3$, and P-P =2.24 \r{A}, P-Se = 2.22 \r{A} for MnPSe$_3$. (b) Spin densities in one unit cell, presenting the antiferromagnetic configuration (the isosurface of 0.4 e/\r{A}$^{3}$). Left: up spin; right: down spin.
}
\end{figure}

Then we use density functional theory \cite{KS65} calculations within the generalized gradient approximation (GGA) \cite{Langreth81,Perdew98} to investigate the basic electronic structure of monolayer MnPX$_3$. The projector-augmented wave potentials are used, as implemented in the Vienna Ab initio simulation package. \cite{Kresse96, Kresse99} A planewave cutoff of 600 eV and a Monkhorst-Pack \textbf{k}-point mesh of $30\times30\times1$ per reciprocal unit cell are adopted. Vacuum slabs at least 15 \r{A} thick are inserted between monolayer MnPX$_3$ to minimized interaction between periodic images. Structure optimizations are performed with a convergence threshold of 0.01 eV/\r{A} on the interatomic forces. To account for the magnetic structure of the divalent transition metal Mn, we use a GGA+\textit{U} approach to describe the on-site electron-electron Coulomb repulsion.\cite{Dudarev98} The value of the isotropic \textit{U} is set to 5 eV, as suggested by a previous assessment of this parameter for the divalent Mn$^{2+}$. \cite{Franchini07} As a calibration for the choice of \textit{U}, the band gap of 2.4 eV is obtained for bulk MnPS$_3$, to be compared with the experimental band gap of bulk MnPS$_3$ of 2.7 eV. \cite{Jeevanandam99} For both bulk and monolayer MnPX$_3$, the antiferromagnetic order of MnPX$_3$ is indeed the more stable in our calculations, compared to the non-magnetic and ferromagnetic states by at least 45 meV per a unit cell. The magnetic moment on each Mn is computed to be about 4.6 Bohr magnetons, in good agreement with the experiments. \cite{Ressouche10} Therefore, the choice of \textit{U} is reasonable and used in all calculations. We compute the interband transition matrix elements using density functional theory at the linear response level. \cite{Gajdos06,Cao12}

The spin density isosurfaces for MnPS$_3$ are displayed in Fig. 4b. The spin densities for the up and down spins show little difference on the {[}P$_2$X$_6${]}$^{4-}$ framework. But the densities for two spins on Mn are indeed well separated and localized on the two Mn ions, providing the expected spin-contrasting asymmetric potential (Fig. 1b). The band structures of monolayer MnPX$_3$ are shown in Fig. 5a. We observe that the band widths near the band gap are quite narrow for MnPX$_3$, between 0.17 and 0.63 eV. The ratio of $U$ to the band widths (a measure of $U/t$) for MnPX$_3$ is greater than 7, indicating that the afm insulating state is expected to be stable in the monolayer limit. \cite{Meng10} MnPS$_3$ and MnPSe$_3$ both show direct band gaps at ${\bf K}_{\pm}$ (2.53 and 1.87 eV, respectively), which fall well within the optical range. This is advantageous for the realization of optical polarization of charge carriers. Within the collinear treatment of magnetism, the Bloch states of two spins are degenerate everywhere in the momentum space. When treating the spins in the non-collinear formalism to account for possible spin-orbit (or spin-valley) interactions, the spins remain degenerate, but band gaps become renormalized as expected. For MnPS$_3$, the gap difference between ${\bf K}_{\pm}$ is negligibly small. But for MnPSe$_3$ the gaps at ${\bf K}_{\pm}$ differ by up to $\sim43$ meV, offering a window for realizing the SHE with equilibrium n- or p-doping (\textit{c.f.} Fig. 2b).

To assess the optical selectivity of valleys by circularly polarized light, we compute the spin-dependent degree of circular polarization, $\eta^{(s)}({\bf k})$,\cite{Yao08,Cao12} between the top valence bands and bottom of conduction bands. This quantity is computed using the density-functional linear response approach, defined as, 
\begin{equation}
\eta^{(s)} ({\bf k}) = \frac{|P^{(s)}_+({\bf k})|^2 - |P^{(s)}_-({\bf k})|^2}{|P^{(s)}_+({\bf k})|^2 + |P^{(s)}_-({\bf k})|^2},
\end{equation}

\noindent where $P^{(s)}_\pm({\bf k})$ are the interband matrix elements of, respectively, left- and right-polarized radiation fields for spin $s$ at $\bf k$, defined for a vertical transition from band $n$ to band $n'$, as $P^{(s)}_\pm({\bf k}; n,n') = \langle n'{\bf k},s| p_x \pm i p_y | n{\bf k},s\rangle$, assuming spin flip is absent in the optical processes. The value of $\eta^{(s)}({\bf k})$ quantifies the relative absorption rates of left- and right-handed photon. As shown in Fig. 5b, monolayer MnPX$_3$ is computed to have perfect circular dichroism at ${\bf K}_{\pm}$. For one spin component they show the valley-selective circular dichroism as previously found for MoS$_2$, \cite{Cao12} and for opposite spins they show a selectivity $\eta^{(s)}(\tau)=2s\tau$, entirely concordant with our theoretical model. The computed selectivity decays as we move away from the high-symmetry ${\bf K}_{\pm}$, but yet substantial polarization is achievable owing to the sizable regions of non-vanishing selectivity.

\begin{figure}[h]
\centering{}\includegraphics[width=6cm]{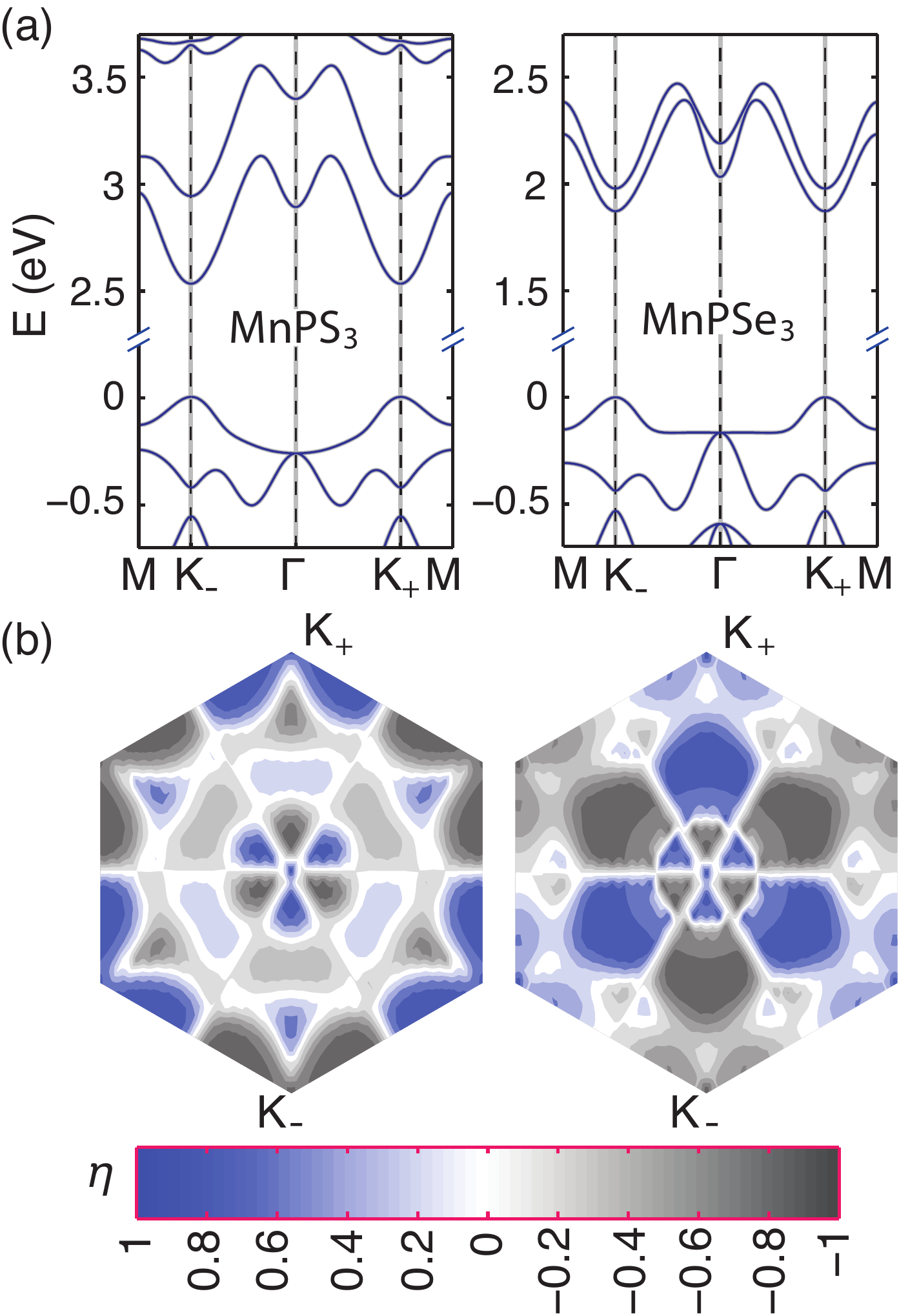}
\caption{Electronic structure of manganese chalcogenophosphates from DFT calculations. (a) The band structures of MnPX$_3$ (X = S, Se) near the band gaps. (b) The momentum-resolved degrees of circular polarization of  MnPX$_3$ (X = S, Se), $\eta^{(s)}({\bf k})$, between to the top of valence bands and the bottom of conduction bands. Only the values of one spin is presented, as in our calculations the other spin takes values equal in magnitude but with opposite signs over the Brillouin zone. At $\Gamma$, the computed optical selectivity is non-zero. This is a numerical artifact because of indeterminacy in $\eta^{(s)}({\bf k})$ in the presence of degeneracy (apart from the spin degeneracy) in the valance bands (see (a)).}
\end{figure}

The optical selectivity of valley interband transitions has been discussed in earlier papers (see, e.g., refs. \cite{Yao08,Cao12,Xiao12}). Here, we phrase it in the language of group theory. Central to the selectivity is the rotational symmetry, $\hat{C}_3$, and the the associated pure rotational group, $C_3$. The point group of a staggered honeycomb lattice (A and B sites are inequivalent) $C_{3v}$. The symmetry at $\bf{K}_\pm$ is, however, $C_3$; i.e., this is an Abelian group that does not allow degeneracy (apart from spin degeneracy). As band degeneracy is absent, then when we operate $\hat{C}_3$ on an eigenstate at a ${\bf k}$ that respects $C_3$ rotational symmetry, we have $\hat{C}_3|n {\bf k}\rangle = \exp (i\varphi_{n{\bf k}}) | n{\bf k}\rangle$, where $\varphi_{n\bf k}=l_{n\bf k}2\pi/3$. The azimuthal quantum number, $l_{n{\bf k}}$, is an integer determined up to modulo 3.  Consider the following transform,

\begin{equation}
\langle n'{\bf k} |\hat{C}_3^{-1}\hat{C}_3 (p_x \pm i p_y)\hat{C}_3^{-1}\hat{C}_3| n{\bf k}\rangle 
=e^{i(\varphi_{n\bf k} - \varphi_{n' \bf k} \pm \frac{2\pi}{3})}P_\pm
\end{equation}

\noindent Clearly, the optical selection rule is modulo$(l_{n\bf k} -l_{n'\bf k},3) = \pm 1$ for left- and right-polarized radiation fields, respectively. This is a mere re-statement of the conservation of angular momentum in the absorption of a single photon absent spin flip. The azimuthal quantum number of the Bloch state, $l_{n\bf k} = l_L + l_M$, which is the sum of two terms: (1) $l_L$ arising from lattice phase winding of the Bloch phase factor, $e^{i{\bf k}\cdot{\bf r}}$, \cite{Yao08} and (2) $l_M = m_l$, where $m_l$ is the magnetic quantum number of the local atomic basis (or, Wannier-like functions) that contribute an additional phase under $\hat{C}_3$. The effect of the local orbital symmetry and orbital ordering at the two valleys has an important role in determining the selection rules, as demonstrated in previous work.\cite{Cao12}

We would also like to remark on how spin-valley optical selection rule develops as a consequence of symmetry breaking in a Fermionic honeycomb lattice. The three-fold rotational invariance of the Hamiltonian implies $P^{(s)}_\pm({\bf K})=0$ for either left- or right-handed ($\pm$) polarization field, when conduction and valence bands are separately non-degenerate (this is not the spin degeneracy, and is ensured when the point group is Abelian).\cite{Cao12} The antiferromagnetic Hamiltonian of Eq. (1) violates both time-reversal symmetry and parity, but accommodates the joint operation, $\hat{\mathcal O}\equiv\hat{\mathcal T}\hat{\mathcal P}$; that is, $[\mathcal{H}^{(s\tau)},\hat{\mathcal{O}}]=0$. When operating on a Bloch state, $\hat{\mathcal{O}}$ preserves its wavevector but inverts spin. Consequently, the Bloch state of the same ${\bf k}$ and opposite spin are degenerate. Moreover, $\hat{\mathcal{O}}$ changes the chirality of light, which implies $P ^{(s)}_\pm({\bf K}_{\tau})= P^{(-s)}_\mp({\bf K}_{\tau})$, to within a phase factor. This is concordant with the notion that the valley-spin index $s\cdot\tau$ constitutes a degree of freedom, which defines chiro-optical selectivity (as well as orbital moments). 

Of note, the proposed optical selectivity may be a convenient assay for antiferromagnetic order in an extremely thin sample, e.g., a monolayer or few-layer sample. This is especially relevant to the recent interest in the yet-to-be-uncovered spin liquid phase on a honeycomb lattice in the intermediate $U$ regime. \cite{Meng10}  The neutron scattering technique commonly used to determine the antiferromagnetic order will become impractical for thin samples. The circularly polarized photoluminescence, on the contrary, works well for these atomically thin samples with sufficient sensitivity.\cite{Cao12,Mak12,Zeng12} In the spin liquid phase, ground-state fluctuations are incessant at $T=0$ K, whereby obliterating the afm order and, hence, the optical selectivity will be absent. Entrance into the symmetry-broken afm ordered r$\acute{\text{e}}$gime, on the other hand, is accompanied by the proposed optical selectivity.

A non-trivial variation of the proposed Hamiltonian of Eq. (1) occurs when the spin-orbit interaction becomes overwhelming, compared to the afm order, $m$. Clearly, when $|\delta| = m$ (see Fig. 1c), the band gap closes in one of the valleys. When $|\delta| > m$, band inversion of one valley will drive the system into a $Z_2$ topological quantum spin Hall state. \cite{Kane05a,Kane05b} In the limit $m=0$, the model reduces exactly to that of Kane and Mele \cite{Kane05a}, which affords time-reversal symmetry-protected edge states. However, when $m$ is finite but smaller than $|\delta|$, time-reversal is lost. The edge states of a ribbon sample will be connected by the $\hat{\mathcal{O}}$-symmetry. This topic and the strategy for tuning the ratio $|\delta|/m$ in real materials are issues worth further pursuit.

We also note that bilayer or few-layer MnPX$_3$ can also have similar optical selectivity, similar to spontaneous symmetry-broken few-layer graphene system.\cite{Fan11} Different kinds of stacking and magnetic order of bilayer MnPX$_3$ (X = S, Se) with intra-layer antiferro- and ferro-magnetism are evaluated in our calculations, where van der waals corrections within GGA are included by optB86b-vdwDF method.\cite{klime10,klime11} Two AA-stacked bilayers with intra-layer antiferromagnetism are nearly degenerate in energy, and more stable than other stacking and magnetic orders. They differ by the inter-layer magnetic configuration; that is, the interlayer couplings are,  respectively, ferromagnetic and antiferromagnetic. By computing the momentum-resolved spin dependent degree of circular polarization, we find that the $s\cdot\tau$-CD is present in ferromagnetically coupled bilayer, which has no inversion center in the magnetic space group. The chiral optical selection rule is absent in the case of antiferromagnetic interlayer coupling, as expected. As the perfect spin-dependent optical selectivity of both monolayer and ferromagnetically coupled bilayer MnPX$_3$, which attests to the emergent electronic degree of freedom presented in the foregoing model, it is therefore of interest to experimentally study these materials in monolayer and few-layer forms, to characterize their optoelectronic and transport behaviors and to explore its potential for application in novel operating paradigms for advanced electronics.

\noindent {\em Acknowledgements}.
We thank the financial support from National Science Foundation of China (NSFC Project 11174009) and China 973 Program (Project 2013CB921900). We are grateful to Prof. Shuang Jia for useful discussions.

\end{document}